\documentclass[twocolumn,superscriptaddress,nofootinbib, floatfix, amsfonts, prd]{revtex4-1}
\usepackage{amsmath}
\usepackage{bm}
\usepackage{graphicx}
\usepackage{subfigure}
\usepackage{float}
\usepackage{mathrsfs}
\usepackage{footmisc}
\usepackage{multirow}
\usepackage{graphicx}
\usepackage{booktabs, ctable}
\usepackage{soul,xcolor}
\usepackage{xcolor, ulem, color, framed}
\usepackage{amssymb}
\usepackage{setspace}
\usepackage[colorlinks, linkcolor=red,anchorcolor=green,citecolor=blue]{hyperref}
\newcommand{\tcc}{$T_{cc}^+$ }
\newcommand{\rms}{$\sqrt{\langle r^2\rangle}$ }

\newcommand{\sblue}{\textcolor{black}}

\graphicspath{{./figs/}}

\begin{document}
\title{Production of doubly charmed hadron $\Xi_{cc}^{++}$ and $T_{cc}^{+}$ in relativistic heavy ion collisions}

\author{Baoyi Chen}\email{baoyi.chen@tju.edu.cn}
\affiliation{Department of Physics, Tianjin University, Tianjin 300350, China}

\author{Meimei Yang}\email{yang\_mm7@tju.edu.cn}
\affiliation{Department of Physics, Tianjin University, Tianjin 300350, China}

\author{Ge Chen}\email{chen\_ge2001@163.com}
\affiliation{Department of physics and astronomy, UCL, Gower St, London WC1E 6BT, England}

\author{Jiaxing Zhao}\email{jzhao@subatech.in2p3.fr}
\affiliation{SUBATECH, Universit\'e de Nantes, IMT Atlantique, IN2P3/CNRS, 4 rue Alfred Kastler, 44307 Nantes cedex 3, France}

\author{Xiao-hai Liu}\email{xiaohai.liu@tju.edu.cn}
\affiliation{Department of Physics, Tianjin University, Tianjin 300350, China}

\date{\today}

\begin{abstract}
Heavy ion collisions provide a unique opportunity for studying the properties of exotic hadrons with two charm quarks. The production of $T_{cc}^+$ is significantly enhanced in nuclear collisions compared to proton-proton collisions due to the creation of multiple charm pairs.
In this study, we employ the Langevin equation in combination with the Instantaneous Coalescence Model (LICM) to investigate the production of $T_{cc}^+$ and $\Xi_{cc}^{++}$ which consists of two charm quarks. We consider $T_{cc}^+$ as molecular states composed of $D$ and $D^*$ mesons. The Langevin equation is used to calculate the energy loss of charm quarks and $D$ mesons in the hot medium. The hadronization process, where charm quarks transform into each $D$ state as constituents of $T_{cc}^+$ production, is described using the coalescence model. The coalescence probability between $D$ and $D^*$ is determined by the Wigner function, which encodes the information of the $T_{cc}^+$ wave function.
Our results show that the $T_{cc}^+$ production varies by approximately one order of magnitude when different widths in the Wigner function, representing distinct binding energies of $T_{cc}^+$, are considered. This variation offers valuable insights into the nature of $T_{cc}^+$ through the analysis of its wave function. The $\Xi_{cc}^{++}$ is treated as a hadronic state produced at the hadronization of the deconfined matter. Its production is also calculated as a comparison with the molecular state $T_{cc}^+$. 
\end{abstract}
\maketitle

\section{introduction}

 The understanding of hadron properties is deeply rooted in nonperturbative Quantum Chromodynamics (QCD), which governs the confinement of fundamental quarks and gluons. In the field of particle physics, the investigation of exotic hadrons has gained significant importance. Exotic hadrons challenge our understanding of the quark model and provide insights into the strong interaction dynamics within QCD~\cite{Chen:2016qju,Liu:2019zoy,Dong:2017gaw,Navarra:2007yw,Braaten:2020nwp,Eichten:2017ffp,Zhao:2020nwy,Huang:2020dci}.
Recently, the LHCb collaboration made a remarkable observation of a doubly charmed state $T_{cc}^+$~\cite{LHCb:2021vvq,LHCb:2021auc}. This discovery sparked debates regarding the nature of multi-charm states, such as X(3872) and $T_{cc}^+$, which are believed to be either compact tetraquark states or loosely bound hadronic molecules. Some theoretical models even propose that these states may arise solely from specific kinetic phenomena~\cite{Guo:2017jvc}. While the study of exotic states has been extensively pursued in particle physics over the past decades~\cite{CDF:2003cab,Bignamini:2009sk,Brambilla:2019esw,Braaten:2022elw,Cho:2019syk}, their investigation in the context of heavy-ion collisions has recently garnered increased attention~\cite{Esposito:2020ywk,Braaten:2020iqw,Chen:2021akx,Yun:2022evm,Hu:2021gdg,Wu:2020zbx}.

In the nucleus-nucleus collisions at the Large Hadron Collider (LHC), the presence of a charm-rich deconfined medium offers a unique environment for studying the production of charmonium and other charmed states~\cite{Andronic:2003zv,Du:2017qkv,Yan:2006ve,Chen:2017duy,Chen:2021uar,Wu:2023djn,Zhao:2017gpq,Zhao:2016ccp,Minissale:2023dct}. After the production from initial parton hard scatterings, these multi-charm bound states are nearly melted in the extremely hot medium due to color screening and parton scatterings with thermal partons~\cite{Satz:2005hx,Brambilla:2008cx,Wen:2022yjx}. Consequently, the majority of experimentally measured multi-charm particles are produced through the coalescence of charm quarks around the phase transition of the medium. 
\sblue{The coalescence model has been employed to study the loosely bound objects such as light nuclei in heavy ion collisions~\cite{Gosset:1976cy,STAR:2001pbk,STAR:2010gyg,Braun-Munzinger:2018hat}. In the loosely bound states with a long formation time, coalescence models work as an effective model and neglect the details of the formation process but only consider the features that the production of the formed states depends on their Wigner function and also the densities of the ingredient particles. The following interactions such as the Coulomb interaction between the formed states and the medium are also usually neglected, or this effect can be partially considered by assuming a later formation of the state. 
}

The $T_{cc}^+$ state, with a mass very close to the $D^+D^{*0}$ and $D^0D^{*+}$ thresholds~\cite{LHCb:2021vvq,LHCb:2021auc}, is regarded as a hadronic molecule with a weak binding energy in this study. Within this framework, charm quarks undergo energy loss in the deconfined medium and hadronize into $D$ mesons during the hot medium's hadronization process. Subsequently, $D$ mesons experience diffusion in the hadronic medium and can potentially combine with another $D^*$ to form a weakly bound molecule. The coalescence probability between two $D$ mesons depends on their relative momentum, distance, and the wave function of $T_{cc}^+$~\cite{Greco:2003vf,Cho:2019lxb,Rapp:2009my,ExHIC:2010gcb,Andronic:2019wva}. This unique situation provides an opportunity to investigate properties related to the wave function of $T_{cc}^+$, which is crucial for understanding the nature of exotic particles. Moreover, the charm-rich environment significantly enhances the production of multi-charm states, increasing the anticipation for experimental measurements in heavy-ion collisions. As coalescence model have been widely used to study the particle production including hadronic molecules in heavy-ion collisions~\cite{ExHIC:2011say,ExHIC:2017smd,Abreu:2022lfy}, we employ the Langevin equation plus coalescence model to study the energy loss of heavy quarks in the medium, and also the production of multi-charm states, like the hadronic state $\Xi_{cc}^{++}$ and the molecular state $T_{cc}^+$. As both $\Xi_{cc}^{++}$ and $T_{cc}^+$ consists of two charm quarks, the difference between their production mainly comes from the different internal potential of the states.

The manuscript is structured as follows. Section II presents the dynamical equations governing the charm energy loss and the coalescence formula for the production of \tcc and $\Xi_{cc}^{++}$. Additionally, a brief overview of the evolution of the hot medium is provided. In Section III, the production of \tcc and $\Xi_{cc}^{++}$ in heavy-ion collisions is plotted and thoroughly analyzed under various configurations. Finally, a conclusion is given in Section IV.

\section{theoretical model}

Heavy quarks have been found to exhibit a strong coupling with the deconfined medium that is generated in relativistic heavy ion collisions~\cite{He:2012df,Banerjee:2011ra,Ding:2015ona}. By assuming small momentum transfer during each scattering event between charm quarks and thermal partons, it is possible to treat the trajectories of charm quarks in the medium as Brownian motion, a phenomenon that has been extensively studied using the Langevin equation~\cite{He:2011qa,Cao:2015hia,Chen:2017duy}.
Previous studies have indicated that heavy quarks experience energy loss in the Quark-Gluon Plasma (QGP) through two distinct mechanisms. At low transverse momentum ($p_T$), the dominant contribution to energy loss for heavy quarks arises from elastic scatterings with thermal partons. Conversely, at high $p_T$, the energy loss of heavy quarks is primarily governed by medium-induced gluon radiation~\cite{Gyulassy:2000fs,Guo:2000nz,Wiedemann:2000tf,Zhang:2003wk,Qin:2007rn,Bass:2008rv,Armesto:2011ht,Majumder:2010qh}.
These two effects can be appropriately incorporated into the Langevin equation, which has been widely employed to investigate the dynamics of heavy quarks in the deconfined medium,
\begin{align}
\label{lan-gluon}
{d{\bf p}\over dt}= -\eta(p) {\bf p} +{\bf \xi} + {\bf f}_g,
\end{align}
where ${\bf p}$ is the three momenta of charm quarks. $\eta(p)$ is the drag coefficient. The momentum diffusion coefficient $\kappa$ is determined through the fluctuation-dissipation relation, $\eta(p)=\kappa/(2TE)$. Here, $T$ and $E=\sqrt{m_c^2+{p}^2}$ represent the medium temperatures and the energy of the charm quark, where the charm quark mass is defined as $m_c=1.5$ GeV. The value of $\kappa$ is determined by the spatial diffusion coefficient, $\kappa \mathcal{D}_s=2T^2$. Previous studies have extracted a value of $\mathcal{D}_s(2\pi T)$ to be approximately $5\sim 7$ in the QGP and larger in the hadronic medium~\cite{He:2012df,Rapp:2018qla}. In this study, we adopt the values of $\mathcal{D}_s(2\pi T)=5$ in the QGP. 
\sblue{In the hadronic medium, $\mathcal{D}_s(2\pi T)$ also varies with temperature ranging from $120$ to $170$ MeV. We approximate this variation by estimating a midpoint value of $8$}, like the cases in previous studies~\cite{Chen:2021akx,Chen:2021uar,Cao:2015hia}.
Assuming the random term ${\bf \xi}$ to be white noise, it satisfies the following correlation:
\begin{align}
\langle \xi^{i}(t)\xi^{j}(t^\prime)\rangle =\kappa \delta ^{ij}\delta(t-t^\prime), 
\end{align}
where $i,j=(1,2,3)$ represent three dimensions. For simplicity, the momentum dependence in the white noise term has been neglected. Besides random elastic collisions, heavy quarks also lose energy via gluon radiation. The force due to the gluon emission is defined as ${\bf f}_g=-d{\bf p}_g/dt$ with the gluon momentum ${\bf p}_g$. 
\sblue{This contribution is absent for hadrons in the hadronic medium.} When the time step is sufficiently small, the number of emitted gluons in the interval $t\sim t+dt$ can be considered as an emission probability~\cite{Cao:2015hia,Zhang:2003wk},
\begin{align}
\label{gluon-spec}
P_{\rm rad}(t,d t) = \langle N_g(t, dt)\rangle = d t \int dx d k_T^2
{dN_g\over dx dk_T^2dt}, 
\end{align}
here, $k_T$ represents the transverse momentum of the emitted gluon. The quantity $P_{\rm rad}$ can also be interpreted as the probability of emitting a single gluon within a time interval, with a value less than unity. Additionally, $x$ denotes the ratio of the gluon energy to the heavy quark energy. The spectrum of emitted gluons from massive quarks, $dN_g/dxdk_T^2dt$, is from the higher-twist calculation for the medium-induced
gluon radiation in perturbative QCD~\cite{Cao:2015hia}.

\sblue{Due to the large mass, heavy quarks are mainly produced via the initial parton hard scatterings, Therefore, the initial spatial distribution of heavy quarks is proportional to the density of nucleon binary collisions, which is the product of the two nuclear thickness function, $dN_{c\bar c}/d{\bf x}_T\propto T_A({\bf x}_T + {\bf b}/2)T_B({\bf x}_T - {\bf b}/2)$. $T_A$ and $T_B$ are the thickness functions of two nuclei. ${\bf b}$ is the impact parameter. The initial momentum distribution of charm quarks is generated by the fixed-order plus next-to-leading log formula (FONLL)~\cite{Cacciari:1998it,Cacciari:2012ny}. Then the initial position and momentum of heavy quarks are generated via the Monte Carlo methods based on the above spatial and momentum distributions. }

During the QGP phase transition at the critical temperature, charm quarks undergo hadronization, resulting in the formation of $D$ mesons. For small transverse momentum ($p_T$), the hadronization process is described by the coalescence model, where charm quarks combine with thermal light quarks to form $D$ mesons. On the other hand, at high $p_T$, the fragmentation process dominates the production of $D$ mesons, with charm quarks emitting gluons. Since our focus is primarily on the total production of \tcc, mainly at low and intermediate $p_T$, we assume that all charm quarks hadronize into $D$ mesons via the coalescence process. This approximation is valid for low and intermediate $p_T$. 
While for $\Xi_{cc}^{++}$, it consists of two charm quarks and one light quark, which is produced near the hadronization hypersurface~\cite{Zhao:2016ccp}. The probability of heavy and light quarks turning into mesons and baryons is written as~\cite{Zhao:2020wcd}:
\begin{align}
\label{eq-Dcoal}
\mathcal{P}_{c+\bar q\rightarrow D}({\bf p}_D,{\bf x}_D)
=&\mathcal{H}_{c\to D}\int {d{\bf p}_c d{\bf x}_c\over (2\pi)^3}  {d{\bf p}_{\bar q}\over (2\pi)^3}
{dN_c\over d{\bf p}_c d{\bf x}_c} {dN_{\bar q}\over d{\bf p}_{\bar q} }
\nonumber \\
\times& 
f_D^W({\bf p}_c, {\bf p}_{\bar q})
\delta^{(3)}({\bf p}_D -{\bf p}_c-{\bf p}_{\bar q}), \nonumber \\
\times& 
\delta^{(3)}({\bf x}_D -{\bf x}_c), \\
\label{eq-Xicoal}
\mathcal{P}_{c+c+q\rightarrow \Xi_{cc}}({\bf p}_\Xi, {\bf x}_\Xi)
=&{g}_{\Xi}\int {d{\bf p}_{c1}d{\bf x}_{c_1}\over (2\pi)^3} {d{\bf p}_{c_2}d{\bf x}_{c_2}\over (2\pi)^3} {d{\bf p}_{q}\over (2\pi)^3}\nonumber\\
\times& {dN_c\over d{\bf x}_{c_1}d{\bf p}_{c_1}}{dN_c\over d{\bf x}_{c_2}d{\bf p}_{c_2}}  {dN_{q}\over d{\bf p}_{q} }
\nonumber \\ \times &
f_{\Xi_{cc}^{++}}^W({\bf p}_{c_1}, {\bf p}_{c_2}, {\bf p}_q,{\bf x}_{c_1}, {\bf x}_{c_2})\nonumber 
\\ \times&
\delta^{(3)}({\bf p}_{\Xi}-{\bf p}_{c_1} -{\bf p}_{c_2}-{\bf p}_{q}) \nonumber \\
 \times& \delta^{(3)}({\bf x}_{\Xi}-{m_{c_1}{\bf x}_{c_1} +m_{c_2q}{\bf x}_{c_2}\over m_{c_1}+m_{c_2q}}),
\end{align}
\sblue{where the position ${\bf x}_D$ of $D$ meson is approximated as the position of charm quarks due to its large mass. 
${\bf p}_c$ and ${\bf p}_{q}({\bf p}_{\bar q})$ are the 
the momentum of the charm and light (anti-light) quarks. The momentum conservation in the reaction is ensured by the presence of the $\delta$ function. So, the produced hadron is not on-shell. The on-shell process is accompanied by emitting other particles, such as photon, $\pi$, which is neglected firstly in this calculation. The quantity $dN_c/d{\bf p}_c d{\bf x}_c$ represents the distribution of charm quarks near the hadronization hypersurface, which is given by the Langevin equation. $dN_{q}/d{\bf p}_{q}$ (or $dN_{\bar q}/d{\bf p}_{\bar q}$) corresponds to the density of thermal light (anti-light) quarks. In our calculation, the latter is usually modeled using a thermal distribution. In Eq.~\eqref{eq-Dcoal}, the spatial part of the Wigner function has been integrated, since light quarks are abundant around charm quarks in QGP which makes spatial conditions of the coalescence process can always be satisfied. So, the Wigner function is only momentum-dependent, $f_D^W({\bf p}_c, {\bf p}_{\bar q})=A \exp(-\sigma^2 {\bf q}_r^2)$, where ${\bf q}_r=(E_c^{cm}{\bf p}_c - E_{\bar q}^{cm}{\bf p}_{\bar q})/(E_c^{cm}+E_{\bar q}^{cm})$ is the relative momentum between the charm and light quark in their center-of-mass frame. $E_{c,\bar q}^{cm}$ are the energy of two particles in their center-of-mass frame. The Gaussian width $\sigma$ is related to the root-mean-square radius of the $D$ meson through $\sigma^2= {4\over 3}{(m_c+m_q)^2\over m_c^2 +m_q^2}\langle r^2\rangle_D $, with $\sqrt{\langle r^2\rangle_D}=0.43$ fm~\cite{Greco:2003vf,Zhao:2020jqu}. The light quark mass is set to be $m=0.3$ GeV. $A$ is a constant factor to make sure the coalescence probability equals 1 when the momentum of charm quark towards zero, ${\bf p}_c\to0$. We force almost all charm quarks to form $D$ mesons. Different from the $\Xi_{cc}^{++}$ as shown in Eq.~\eqref{eq-Xicoal}, We use 
Eq.~\eqref{eq-Dcoal} to describe the coalescence process of $D$ where we don't use the degeneracy factor for different $D$ meson states instead a hadronization fraction $\mathcal{H}_{c\to D}$. $\mathcal{H}_{c\to D}$ denotes the hadronization fraction of charm quarks transitioning into different states of $D$ mesons, such as $D^{0,±}$, $D^{*0}$, and $D^{*\pm}$. These values can be estimated from the experimental data and thermal model~\cite{ALICE:2018lyv,ALICE:2021dhb,Zhao:2024ecc}, which are taken as $\mathcal{H}_{c\rightarrow D^0,D^+, D^{*0}, D^{*+}}\approx (11.3\%, 11.3\%, 16.1\%, 16.1\%)$. The rest of the fractions are to form other charmed hadrons, such as $D_s$, $\Lambda_c$, $\Xi_c$, and so on.
The advantage of this way is to avoid considering all charmed hadrons and only focus on $D$ mesons. This will increase the statistic effectively and it's proved the same as the way to consider all charmed hadrons separately~\cite{Chen:2021akx}. 
}

\sblue{For the case of $\Xi_{cc}^{++}$, it consists of two charm quarks and one light quark. The light quark is abundant around the charm quark and with a small mass. In this case, one can assume that the coalescence probability of $\Xi_{cc}^{++}$ mainly depends on the distributions of two charm quarks, and the three-body Wigner function can be simplified as the form of the two-body situation $c-cq$. Because charm quarks are rare particles in QGP, the spatial part of the Wigner function should be considered as well. The Wigner function of $\Xi_{cc}^{++}$ can be expressed as $f_{\Xi_{cc}^{++}}^W=8 \exp(-{\bf x}_r^2/\sigma^2)\exp(-\sigma^2 {\bf q}_r^2)$, ${\bf x}_r={\bf x}_{c_1}-{\bf x}_{c_2q}$ is the relative distance between charm quark ($c_1$) and charm-light-pair ($c_2q$) in their center-of-mass frame, where the position of charm-light-pair is the same as the charm quark, ${\bf x}_{c_2q}={\bf x}_{c_2}$. 
The Gaussian width $\sigma$ is related to the root-mean-square radius of $\Xi_{cc}^{++}$ through $\sigma^2= {4\over 3}{(m_c+m_{cq})^2\over m_c^2 +m_{cq}^2}\langle r^2\rangle_{\Xi_{cc}^{++}} $, $\sqrt{\langle r^2\rangle_{\Xi_{cc}^{++}}}=0.5$ fm~\cite{Zhao:2016ccp} in the following calculations. $m_{cq}=m_c+m_q$ is the mass of charm-light-pair. The spin and color degeneracy factor of  $\Xi_{cc}^{++}$ is extracted to be $g_{\Xi_{cc}^{++}}=1/54$. }

In the hadronic medium, $D$ mesons continue to undergo diffusion until the kinetic freeze-out stage. Since the binding energy of the \tcc molecule is considered to be small, \tcc is produced at temperatures close to the kinetic freeze-out temperature.
We also adopt a Gaussian form for the Wigner function of \tcc, given by $f^W_{T_{cc}^+}=8\exp(-{\bf x}_r^2/\sigma^2)\exp(-\sigma^2 {\bf q}_r^2)$~\cite{Greco:2003vf}. Here, ${\bf x}_r$ and ${\bf q}_r$ denote the relative distance and relative momentum between the $D$ meson and $D^*$ meson, respectively.
To study the production of $T_{cc}^+$ in the hadronic medium, we consider different values for the root-mean-square radius of $T_{cc}^+$, which carries information about the $T_{cc}^+$ wave function.
In nucleus-nucleus collisions, the parton densities of nucleons in the nucleus can be modified by cold nuclear matter effects, such as the shadowing effect. This can also impact the initial distributions of charm quarks. The shadowing factor is calculated using the EPS09 package and is applied directly to the aforementioned initial distribution of charm quarks. Furthermore, the gluons can undergo scattering with other nucleons, resulting in additional energy before fusing into a charm pair. This additional energy will be transferred to the charm quarks and modify their initial momentum distributions. Since charm quarks experience significant energy loss in the quark-gluon plasma, we neglect the modification of the Cronin effect on the initial distribution of charm quarks in this study.

The deconfined matter generated in relativistic heavy-ion collisions exhibits properties similar to those of a nearly perfect fluid~\cite{Song:2010mg,Hirano:2005wx}. The evolution of the hot medium can be effectively simulated using hydrodynamic equations~\cite{Song:2010mg,Schenke:2010rr,Pang:2012he,Shen:2014vra}. We employ the MUSIC package~\cite{Schenke:2010nt,Schenke:2010rr} to simulate the dynamical evolutions of the deconfined medium and the hadronic gas. For the equation of state (EoS) of the medium, we employ the EoS parametrized with the lattice EoS at zero baryon density from the HotQCD Collaboration and the hadron resonance gas EoS~\cite{Bernhard:2016tnd,HotQCD:2014kol}. 
 There is a crossover phase transition between two phases~\cite{Bazavov:2011nk}. We specify a critical temperature $T_c = 170$ MeV, above and below which the medium is treated as deconfined and confined, respectively.
The initial conditions of the hot medium and the starting time of hydrodynamic evolutions are determined based on the final spectrum of light hadrons~\cite{Zhao:2017yhj}.

\section{$T_{cc}$ production in heavy-ion collisions}
As $T_{cc}^+$ is considered to be a bound state of $D$ and $D^{*}$ mesons, its production can be significantly enhanced in nuclear collisions due to the large number of charm quarks involved. Since experiments only measured the $T_{cc}^+$, we will take the isospin to be $0$ and calculate the production of this specific state via the coalescence of $D^{+/0}D^{*0/*+}$. 
The production cross-section of charm pairs has been experimentally measured. In central rapidity proton-proton (pp) collisions at a center-of-mass energy of $\sqrt{s_{NN}}=5.02$ TeV, the rapidity-differential cross-section is $d\sigma_{pp}/dy=1.165$ mb~\cite{ALICE:2021dhb}. The production of \tcc depends on both the phase space distribution of $D$ mesons and the Wigner function of \tcc. The positions and momenta of $D^{0/+},D^{*0/*+}$ before the coalescence of \tcc are determined using the Langevin equation. The Wigner function, which is related to the \tcc wave function, is approximated as a Gaussian function with a width determined by the root-mean-square radius $\sqrt{\langle r^2\rangle}$. Considering that the binding energy of \tcc is very small, it can be easily dissociated by hadronic scatterings in the hadron gas. We assume that the loosely bound molecule is produced at a low temperature close to the kinetic freeze-out temperature of the medium, and take different values of $\sqrt{\langle r^2\rangle}$ into calculation due to the uncertainty of the \tcc wave function.  

\begin{figure}[!hbt]
\centering
\includegraphics[width=0.4\textwidth]{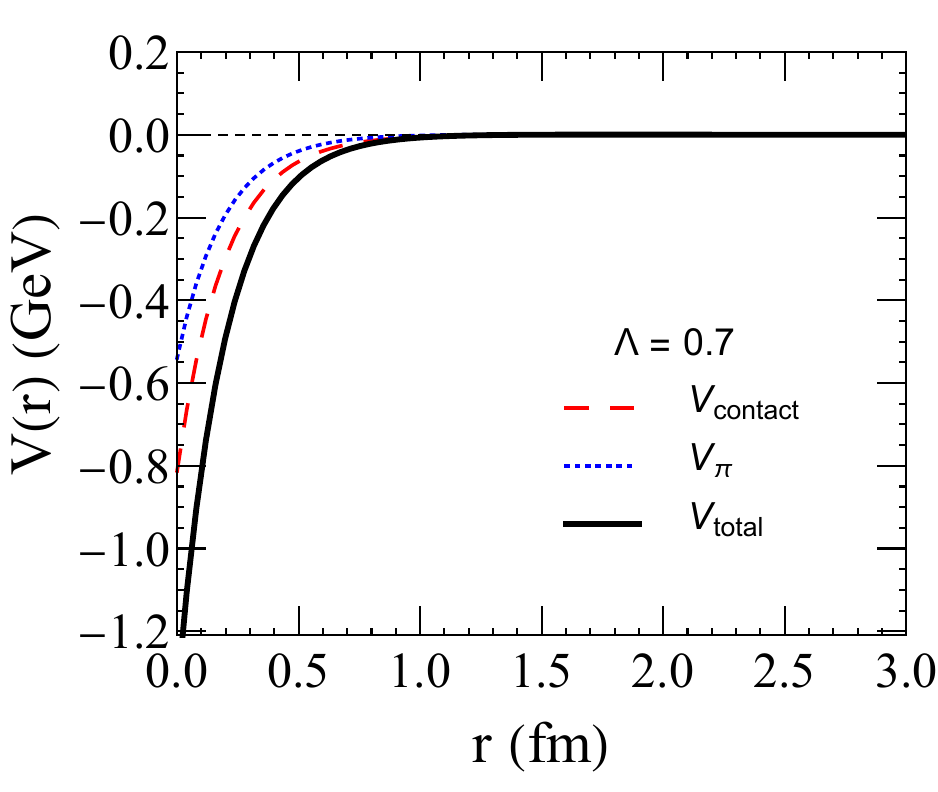}\\
\includegraphics[width=0.4\textwidth]{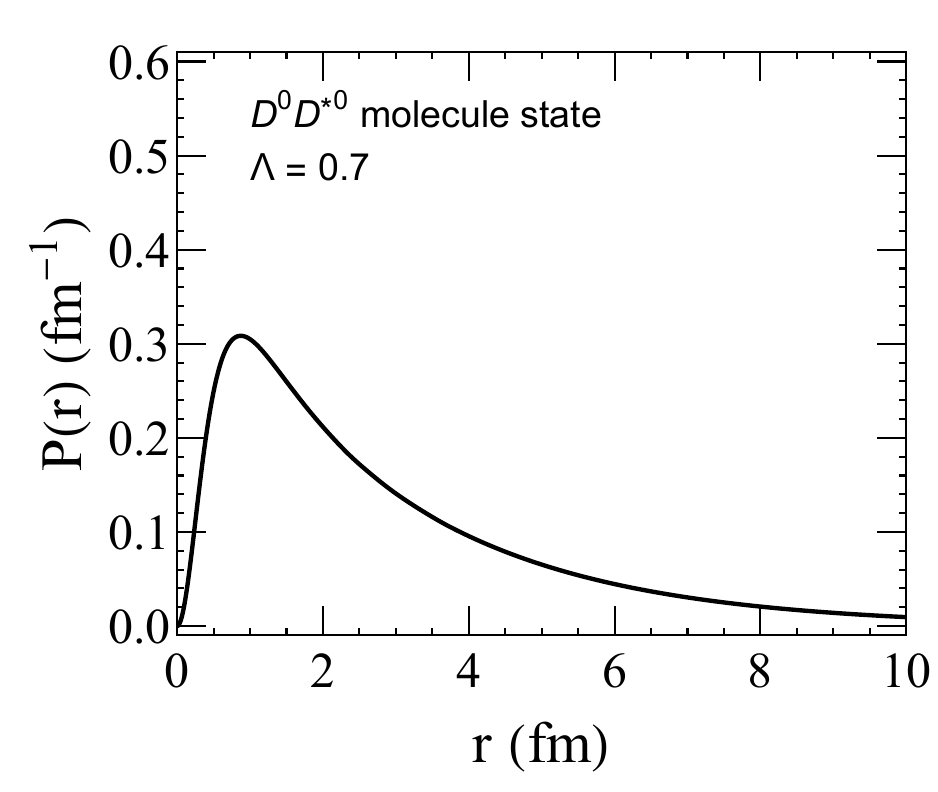}
\caption{(Upper panel)the total potential between $D^0$ and $D^{*0}$. (Lower panel)the radial probability $P(r)=|\psi(r)|^2r^2$ of loosely bound molecular state $T_{cc}^+$.
}
\hspace{-0.1mm}
\label{fig-DDstar}
\end{figure}

To get an estimation of the root-mean-radius of \tcc states, the $DD^*$ potentials are studied within the framework of heavy meson chiral effective field theory.  The effective potentials of the $DD^*$ system from the contact and one-pion exchange (OPE) diagrams at the leading order are considered. The contact terms mainly affect the short-range interaction between particles while the OPE contribution determines the behavior of the long-range interaction. Both the contact and OPE interaction are isospin $I$ dependent, where $I$ is the isospin of the formed molecular state.
The contact terms lead to the attractive interaction in the $I = 0$ channel while repulsive interaction in the $I = 1$ channel.  OPE interaction is attractive in both cases. So, the total potential in the short distance for the $I = 1$ is repulsive but for the $I = 0$ channel is attractive, see detail in Ref.~\cite{Xu:2017tsr}. In this study, we only consider the $I=0$ channel, which has attractive potential and supports the formation of the molecular state. The potential in momentum can be shown as,
\begin{eqnarray}
V_{\rm contact}(q)&=&-2D_a-2D_b+6E_a +6E_b, \nonumber\\
V_{\rm OPE}(q)&=&-{g^2\over 4f_\pi^2}{q^2\over q^2+m_\pi^2}.
\end{eqnarray}
We take the same parameters as~\cite{Xu:2017tsr}, $D_a=-6.62$ $E_a=-5.74$, and $D_b=E_b=0$. The coupling constant $g=0.65$ and $f_\pi=86$ MeV. $m_\pi=135$ MeV is $\pi$ mass.
The potential in coordinate space can be obtained via transformation. 
We introduce the monopole type form factor at each vertex to take into account the size effect of $D^{(*)}$ mesons
\begin{eqnarray}
F(q)=\left(\Lambda^2-m^2 \over \Lambda^2+q^2 \right)^2.
\end{eqnarray}
The cut-off parameter $\Lambda$ is determined from the size of $D^{(*)}$ and is usually taken around 1GeV. Solving two-body Schr\"odinger equation with this potential, we can get the wavefunction and binding energy of $D^0 D^{*0}$ bound state, as shown in TABLE~\ref{tab2} and Fig.~\ref{fig-DDstar}. Our results are consistent with other studies with many other vector meson exchange potential~\cite{Li:2012ss,Ohkoda:2012hv}. 

\begin{table*}[!hbt]
\renewcommand\arraystretch{1.2}
\caption{The binding energy (B.E.) and average radius $\langle r \rangle$ of molecule $T_{cc}^+$ states with different cut-off parameter $\Lambda$ in potential.}
\label{tab2}
\setlength{\tabcolsep}{2.5mm}
\begin{tabular}{c|cccccccc}
	\toprule[1pt]\toprule[1pt] 
	 $\Lambda$&0.66 & 0.68 & 0.7 & 0.75 & 0.8  & 0.9 & 1.0 \tabularnewline
	\midrule[0.6pt]
	\text{B.E.}(keV)           &  0.03 & 0.3 & 0.8 & 2.61   & 5.9  &17.7 & 38.4 \tabularnewline
	$\langle r \rangle$(fm) & 5.04 & 3.87& 2.94& 1.74 & 1.26  & 0.83 & 0.62\tabularnewline
	\bottomrule[1pt]\bottomrule[1pt]
	\end{tabular}
\end{table*}

\begin{figure}[!hbt]
\centering
\includegraphics[width=0.4\textwidth]{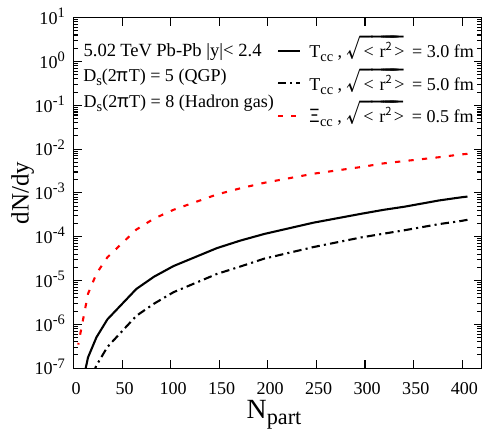}
\caption{ The $T_{cc}^+$ production, measured by the $dN/dy$ as a function of the number of participants $N_{\rm part}$, is presented for central rapidity in Pb-Pb collisions at $\sqrt{s_{NN}}=5.02$ TeV. We select a coalescence temperature of $T_{\rm coal}=0.12$ GeV~\cite{Chen:2021akx,Song:2008si}. Two typical values for the root-mean-square radius of $T_{cc}$, representing different binding strengths, are considered as $\sqrt{\langle r^2\rangle}=(3.0, 5.0)$ fm. The spatial diffusion coefficients for charm quarks and $D$ mesons are chosen as $\mathcal{D}_s(2\pi T)=5$ and $8$ respectively in the QGP and the hadronic medium. The red dashed line represents the hadronic state $\Xi_{cc}^{++}$ as a comparison, which is produced near the hadronization hypersurface of the QGP medium.
}
\label{lab-dNdy}
\end{figure}

Figure \ref{lab-dNdy} displays the production of \tcc and $\Xi_{cc}^{++}$ for various collision centralities. The production of \tcc is directly proportional to the densities of $D$ and $D^*$ mesons, which increase as one moves from peripheral to central collisions. This leads to an enhanced \tcc production compared to that in proton-proton (pp) collisions. We consider \tcc production with different root-mean-square radii, which represent varying binding strengths of \tcc. As the value of the root-mean-square radius decreases, a larger number of $D$ and $D^*$ mesons, with even larger relative momentum, can effectively combine to form the \tcc molecular state. 
The production of \tcc exhibits an increase of approximately 3.5 times when the value of the root-mean-square radius (\rms) varies from 5.0 fm to 3.0 fm, indicating a stronger binding scenario for \tcc. The ultimate production of \tcc is influenced by the selection of \tcc geometry size. This feature renders our model well-suited for extracting information about \tcc through its final production in heavy-ion collisions.
In peripheral collisions, both the volume of the QGP and the number of charm pairs are significantly reduced, resulting in a suppression of \tcc production from the coalescence process. The total production of \tcc is predominantly governed by the primordial production, similar to the case of $J/\psi$. 
We also study the production of $\Xi_{cc}^{++}$ which also consists of two charm quarks and has been regarded as a hadronic state. Due to the strong binding potential, its production becomes around 10 times larger than the molecular \tcc, see the red dashed line in Fig.~\ref{lab-dNdy}.
In our theoretical calculations in Fig. \ref{lab-dNdy}, where we have not considered the primordial production, the \tcc production is underestimated at small $N_{\rm part}$. However, for large $N_{\rm part}$ corresponding to central collisions, the extremely hot medium dissociates almost all of the charmed hadrons, making the coalescence model a suitable approach to describe the production of charmed particles. 

\begin{figure}[!hbt]
\centering
\includegraphics[width=0.4\textwidth]{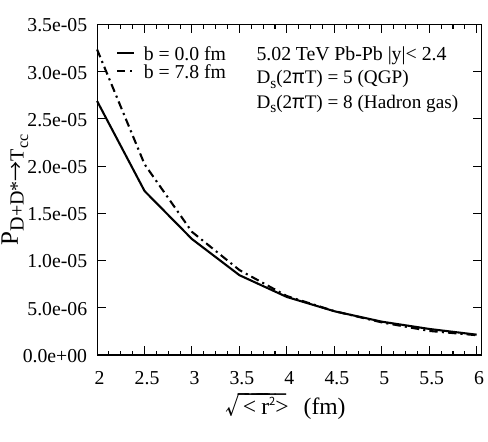}
\caption{ The coalescence probability between one $D$ and $D^*$ to form \tcc varies with the root-mean-square radius $\sqrt{\langle r^2\rangle}$ of \tcc in the central rapidity of $\sqrt{s_{NN}}=5.02$ TeV Pb-Pb collisions. Two collision centralities are selected with the impact parameter to be $b=0$ and $7.8$ fm (minimum bias collisions). The coalescence temperature of the \tcc production is $T_{\rm coal}=0.12$ GeV. 
}
\label{lab-P-varR}
\end{figure}

To investigate the influence of the binding strength of \tcc on its production, we calculate the coalescence probability between one randomly generated $D$ meson and one randomly generated $D^*$ meson in the hot medium. The results are shown in Fig.~\ref{lab-P-varR}. As the value of \rms increases, indicating a weaker binding strength of \tcc, the momentum part of the coalescence probability of $D$ and $D^*$  decreases due to the presence of $\exp(-\sigma^2 q_r^2)$ in the Wigner function. However, the coalescence probability increases in the spatial part, characterized by $\exp(-x_r^2/\sigma^2)$. Two parts are connected, and the combined effect results in a significant suppression of the coalescence probability $P_{D+D^*\rightarrow T_{cc}}$ with the increasing \rms, as demonstrated in Fig.~\ref{lab-P-varR}. We consider two collision centralities, namely the most central collisions and the minimum-bias collisions, which correspond to the production of two different volumes of the QGP. 
When the value of \rms is small, the coalescence probability $P_{D+D^*\rightarrow T_{cc}}$ depends on the diffusions of charm quarks and D mesons in the expanding medium. 
In scenarios where the medium expansion is more intense and persists for a longer duration, such as in the case of $b=0$, $D$ and $D^*$ mesons experience greater diffusion and spread out over a larger volume. 
Consequently, the coalescence probability between a $D$ meson and a $D^*$ meson is reduced when the geometry size of \tcc is smaller. However, as \rms increases, $D$ and $D^*$ mesons that are farther apart can also combine to form \tcc. This feature results in the coalescence probability $P_{D+D^*\rightarrow T_{cc}}$ being less sensitive to the volume of the QGP, as observed in the cases of the QGP generated in the most central (b=0) and minimum-bias collisions, as shown by the two lines in Fig. \ref{lab-P-varR}.

\begin{figure}[!hbt]
\centering
\includegraphics[width=0.4\textwidth]{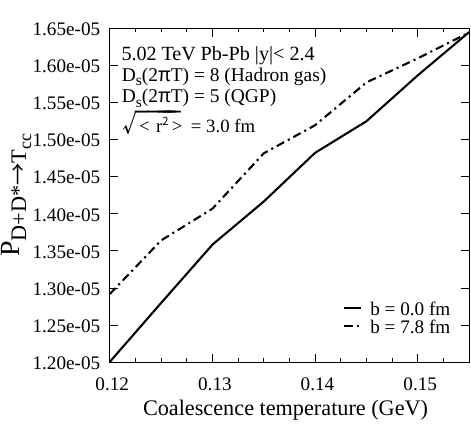}
\caption{ The coalescence probability $P_{D+D^*\rightarrow T_{cc}}$ between one $D$ and one $D^*$ to form \tcc varies with the coalescence temperature of \tcc. The impact parameter is selected as $b=0$ and $b=7.8$ fm respectively, which correspond to the most central collisions and the minimum-bias collisions. The root-mean-square radius used in the Wigner function of \tcc is selected as {$\sqrt{\langle r^2\rangle}=$} 3.0 fm. 
}
\label{lab-dNdy-varT}
\end{figure}

When the binding energies of $T_{cc}^+$ become different, the corresponding coalescence temperature should also differ. In Fig. \ref{lab-dNdy-varT}, we calculated the coalescence probability between one $D$ and $D^*$ meson at different hypersurfaces specified with the coalescence temperature $T_{\rm coal}$. The hot medium is chosen in the most central collisions at 5.02 TeV Pb-Pb collisions. {The root-mean-square radius is $\sqrt{\langle r^2 \rangle} = 3.0$ fm.} When $\sqrt{\langle r^2 \rangle}$ is large (such as $\sqrt{\langle r^2 \rangle} \ge 3$ fm), the value of $P_{D+D^* \rightarrow T_{cc}}$ depends less on the volume of the hot medium at different collision centralities. Therefore, we only consider one collision centrality in Fig. \ref{lab-dNdy-varT}. In the expanding hot medium, when the coalescence temperature varies from 0.12 to 0.16 GeV, the distance between one $D$ and $D^*$ becomes shorter as they move to the regions satisfying the coalescence temperature, which increases the coalescence probability between $D$ and $D^*$. {The value of $P_{D+D^* \rightarrow T_{cc}}$ increases by around 40\% when $T_{\rm coal}$ changes from 0.12 GeV to 0.16 GeV.}

\section{summary}
We utilize the Langevin equation plus coalescence model to investigate the dynamical evolution of charm quarks in the quark-gluon plasma and $D$ mesons in the hadronic medium, as well as the production of \tcc through the combination of $D$ and $D^*$ mesons. The production of \tcc is closely linked to the phase space densities of $D$ mesons at the coalescence hypersurface, which is obtained from the Langevin equation, and the Wigner function, which is related to the wave function of \tcc. To elucidate the properties of \tcc, we consider different binding energies by employing different Wigner functions to calculate the \tcc production. Since the phase space distributions of $D$ mesons are typically determined by fitting the observables of $D$ mesons in heavy-ion collisions, the uncertainty in \tcc production mainly arises from the Wigner function at finite temperatures. Therefore, we vary the width of the Wigner function and the coalescence temperatures to study the production of molecular \tcc in different centralities of $\sqrt{s_{NN}}=5.02$ TeV Pb-Pb collisions. The $\Xi_{cc}^{++}$ is also calculated as a comparison. Its production becomes larger than the molecular \tcc due to the large binding energies of $\Xi_{cc}^{++}$ reflected in the width of the Wigner function. For \tcc, when the wave function becomes broader corresponding to a weaker binding strength, the coalescence probability between $D$ and $D^*$ to form \tcc is reduced evidently. When the 
coalescence temperature increases, \tcc is produced at an earlier stage of the hot medium, where the spatial density of $D$ mesons is higher, thereby enhancing the \tcc production. The production of \tcc is sensitive to the details of the Wigner function. This characteristic makes our model particularly suitable for exploring  the information about \tcc through its final production and investigating the properties of exotic states containing multiple charm quarks.

\vspace{2cm}
{\bf Acknowledgement:} This work is supported by the National Natural Science Foundation of China
(NSFC) under Grants No.12175165, No.11975165 and No.12235018. J. Zhao is supported by the European Union’s Horizon 2020 research and innovation program under grant agreement No. 824093 (STRONG-2020).

\end{document}